\documentclass[aps,pre,reprint,showpacs,superscriptaddress]{revtex4-1}

\usepackage{graphicx}
\usepackage{subfigure}
\usepackage{color}
\usepackage{amsmath}
\usepackage{lipsum}
\usepackage{amssymb}
\usepackage{epstopdf}
\usepackage[linktocpage,colorlinks=true,linkcolor=blue,citecolor=blue,breaklinks=true]{hyperref}
\usepackage{verbatim}
\usepackage{breakcites}
\usepackage[version=3]{mhchem}

\newcommand{\be}{\begin{equation}}
\newcommand{\ee}{\end{equation}}

\DeclareMathOperator\erfc{erfc}
\DeclareMathOperator\erf{erf}

\begin{document}

\preprint{}

\title{Statistical Mechanics and the Climatology of the Arctic Sea Ice Thickness Distribution}

\author{Srikanth Toppaladoddi}
\affiliation{Yale University, New Haven, USA}

\author{J. S. Wettlaufer}
\affiliation{Yale University, New Haven, USA}
\affiliation{Mathematical Institute, University of Oxford, Oxford, UK}
\affiliation{Nordita, Royal Institute of Technology and Stockholm University, SE-10691 Stockholm, Sweden}

\email[]{john.wettlaufer@yale.edu}

\date{\today}

\begin{abstract}
We study the seasonal changes in the thickness distribution of Arctic sea ice, $g(h)$, under climate forcing.  Our analytical and numerical approach is based on a Fokker-Planck equation for $g(h)$ (Toppaladoddi \& Wettlaufer \emph{Phys. Rev. Lett.} {\bf 115}, 148501, 2015), in which the thermodynamic growth rates are determined using observed climatology. In particular, the Fokker-Planck equation is coupled to the observationally consistent thermodynamic model of Eisenman \& Wettlaufer (\emph{Proc. Natl. Acad. Sci. USA} {\bf 106}, pp. 28-32, 2009). We find that due to the combined effects of thermodynamics and mechanics, $g(h)$ spreads during winter and contracts during summer. This behavior is in agreement with recent satellite observations from CryoSat-2 (Kwok \& Cunningham, \emph{Phil. Trans. R. Soc. A} {\bf 373}, 20140157, 2015). Because $g(h)$ is a probability density function, we quantify all of the key moments (e.g., mean thickness, fraction of thin/thick ice, mean albedo, relaxation time scales) as greenhouse-gas radiative forcing, $\Delta F_0$, increases.  The mean ice thickness 
decays exponentially with $\Delta F_0$, but {\em much slower} than do solely thermodynamic models.  This exhibits the crucial role that ice mechanics plays in maintaining the ice cover, by redistributing thin ice to thick ice--far more rapidly than can thermal growth alone.
\end{abstract}


\maketitle

\section{Introduction}
Arctic sea ice is one of the most sensitive components of the Earth's climate system and serves as a bellwether for global scale change. The recent decline in both the areal extent and the average thickness of sea ice, as evidenced by satellite and submarine measurements, drives study of its origins \cite{OneWatt}. The key quantity of interest in the geophysical-scale description of sea ice is its volume; while daily areal extent is routinely measured using satellites, it is a challenge to understand the evolution of the ice volume because of the difficulties involved in the measurement of the thickness, $h$ \cite{OneWatt}. 

To study the evolution of the ice volume, one could treat ice as a continuum and construct the mass, momentum, and energy balance equations \cite{Untersteiner:book}. However, such a description is incomplete without the knowledge of the rheology and physical properties, such as the albedo and thermal growth rate, of the ice pack. These physical properties depend strongly on the thickness.  Thus, this implies that in order to complete a continuum description, one should first determine these properties for the ice pack.

The key step in the construction of such a description was taken in 1975 by Thorndike \emph{et al.} \cite{Thorndike:1975}, who introduced the concept of thickness distribution, $g(h)$. It is defined as follows: Consider a region with area $R$ that is sufficiently large to contain a range ice of different thicknesses. Then the integral
\be
\int_{h_1}^{h_2} g(h) \, dh = \frac{A}{R}
\ee
gives the fraction of that area ($A/R$) that contains ice of thicknesses between $h_1$ and $h_2$, and the dependence of $g(h)$ on space and time is implicit. The spatio-temporal evolution of $g(h)$, subject to wind, thermal and mechanical forcing, is governed by \cite{Thorndike:1975}:
\be
\frac{\partial g}{\partial t} = -\nabla \cdot \left(\boldsymbol{u}  g\right)-\frac{\partial}{\partial h} \left(f g\right) + \psi,
\label{eqn:thorndike}
\ee
where $\boldsymbol{u} $ is the horizontal velocity of ice pack, $f$ is the thermal growth/melt rate of ice, and $\psi$ is the redistribution function that accounts for all the mechanical interactions between ice floes (ridging, rafting, and formation of open water). The principal difficulty in solving equation \ref{eqn:thorndike} came from $\psi$ whose general form could not be deduced from observations. Thorndike \emph{et al.} \cite{Thorndike:1975} separately considered the cases of the formation of open water and pressure ridges, and constructed simple models of $\psi$ for these events based on physical arguments. The general form of $\psi$ was taken to be the combination of the above mentioned cases. Numerical integration of equation \ref{eqn:thorndike} using initial conditions from the limited submarine measurements resulted in $g(h)$'s that were qualitatively similar to those from observations. However, equation \ref{eqn:thorndike} remained intractable due to the lack of a closed mathematical form for $\psi$.   Indeed, Thorndike \emph{et al.} \cite{Thorndike:1975} noted that ``The present theory suffers from a burdensome and arbitrary redistribution function $\psi$.''

Thorndike, in a later study \cite{Thorndike:1992}, made two calculations in order to understand the nature of $\psi$ and its role in the evolution of $g(h)$. In the first calculation he obtained $\psi$ by assuming a steady state and solving for:
\be
\psi = g \, \nabla \cdot \boldsymbol{u} + \frac{\partial}{\partial h} \left(\overline{f} g\right),
\label{eqn:thorndike1992}
\ee
where $\overline{f}$ is the annually averaged thermal growth rate from the one-dimensional thermodynamic model of Maykut \& Untersteiner (MU71) \cite{MU71}, and $g(h)$ was taken  from observations. Depending on the values of $d = \nabla \cdot \boldsymbol{u} $, the solutions displayed the following features: (a) $\psi$ provided a source of open water; (b) ice of thickness less than a certain value $h^*$ was used to build pressure ridges, and hence $\psi$ was a sink for this range of thickness; and (c) $\psi$ was a source of ice thicker than $h^*$. 

For his second calculation, Thorndike formulated the original equation as a Markov process; and by assuming the forms of $\overline{f}$ and $\psi$ he solved for the steady state. In constructing the matrices of $\overline{f}$ and $\psi$ he used the following principle: If ice of initial thickness $h_i$ grew either by thermal growth or by ridging to a final thickness $h_f$, then the process that led to this increase would act as a sink for $g(h=h_i)$ and source for $g(h=h_f)$; similar arguments hold in the case of thinning. Divergence affected ice of all thicknesses, and there was a source term for open water. He assumed that $\psi$ depended on the random short-term strain ${\bf e}$ in the ice. By using different values of $d$ and ${\bf e}$, he was able to show the effects of different processes on $g(h)$. The following is a brief summary of his findings: 
\begin{enumerate}
\item When $d = {\bf e} = 0$, $g(h) = \delta(h-H_{eq})$. Here $\delta(x)$ is the Dirac-delta function and $H_{eq}$ is the ``equilibrium" thickness. For a typical profile, $g(h)$ attains the maximum value at $h=H_{eq}$. 
\item Choosing ${\bf e} > 0$ and $d=0$ leads to a spread in $g(h)$ on both sides of the maximum, but for $d>0$ and ${\bf e} = 0$ the spread is only in the thinner side. 
\item In order to obtain a steady solution, it is necessary for ${\bf e} \neq 0$ when $d < 0$. Thus, the solution in this case has very little thin ice.
\item The solutions with $d = 0$ and ${\bf e} > 0$ qualitatively resemble the observed $g(h)$.
\end{enumerate}
This study considerably improved our understanding of $\psi$, but left the following key issues open: 
\begin{enumerate}
\item A closed form of $\psi$ was still lacking, which prohibited any systematic mathematical analysis of equation \ref{eqn:thorndike}.
\item It was assumed that ice only from a particular range of thickness could ridge to produce thicker ice, but this is generally not the case \cite{VW08}.
\item It was difficult to use this framework to study seasonal changes in $g(h)$.
\end{enumerate}

The theoretical investigation of the evolution of $g(h)$ was complemented by observations of thickness in the central Arctic, which revealed that $g(h) \sim e^{-h/H}$ for thick ice. Thorndike \cite{Thorndike:2000} thus constructed simpler models for the thermal and mechanical processes to explain the observed exponential tail. For the thermal process, he assumed $\overline{f}(h) = F \times (H_{eq}-h)$, where $F^{-1}$ is the time scale required to reach $H_{eq}$. The rate of formation of open water and ridges was assumed to be $r$. 
Using dimensional arguments he related $H$ to $H_{eq}$ by: 
\be
H = \mathcal{G}\left(\frac{F}{r}\right) H_{eq},
\label{eqn:H-Heq}
\ee
where $\mathcal{G}$ is some function of $F/r$. Thorndike \cite{Thorndike:2000} argued that because there are a large number of interacting floes, the larger the fraction of a certain thickness the larger the probability of participation in ridging to produce thicker ice. From this logic he arrived at the following form for $\psi$; 
\be
\psi = r \left[\delta(h) - 2 \, g(h) + \int_0^h g(h') \, g(h-h')\, dh'\right], 
\label{eqn:scm2000}
\ee
where, $\delta(h)$ is the source of open water, $-2g(h)$ is the sink term for the ice that is used for ridging, and the convolution term represents the sum of all interactions that produce ice of thickness $h$. While this approach overcame limitation 2 from the previous study, the nonlinear integro-differential equation could only be solved numerically. The solutions displayed exponential tails, showing that the simple rules for the thermal and mechanical interactions were sufficient to obtain $g(h)$ for thick ice that were in qualitative agreement with the observations. 

Recently, Godlovitch \emph{et al.} \cite{godlovitch2011} generalized Thorndike's approach (equation \ref{eqn:scm2000}) using Smoluchowski coagulation models. These models describe the evolution of a population of particles that can interact in pairs to change their mass, with the rate of coalescence that depends on their mass. Using this formalism, $\psi$ was represented as
\be
\begin{split}
\psi &= C(K,t) \, \delta(h) + \frac{1}{2} \int_0^h K(h',h-h')\, g(h')\, g(h-h') dh' \\
& - \int_0^{\infty} K(h,h')\, g(h)\, g(h')\, dh',
\end{split}
\label{eqn:scm}
\ee
where $K(h,h')$ is the rate kernel and $C(K,t)$ is introduced to ensure that $g(h)$ is normalized. Numerical solutions to equation \ref{eqn:scm} displayed exponential and quasi-exponential tails for a variety of $K(h,h')$, indicating that the nature of the ridging process is not sensitive to the choice of rate kernel. However, the choice of $K(h,h')$ would become important for quantitative comparison with observations \cite{godlovitch2011}.

Finally, the World Climate Research Programme Coupled Model Intercomparison Project Phase 5 (CMIP5) models, which use momentum equations for the ice pack and hence a wide range of constitutive models for the ice rheology, are known to poorly represent the spatial patterns of ice thickness \cite{Stroeve:2014}.   This highlights the potential utility of the  probabilistic approach to understanding the large scale behavior of the ice pack that motivated the original theory of Thorndike \emph{et al.} \cite{Thorndike:1975}.  

\section{A Statistical Mechanics Based Theory \cite{TW2015}}
When studying equation \ref{eqn:thorndike}, it is important to realize that there is a separation of length and time scales over which the mechanical processes (e.g., ridging and rafting) act relative to the evolution of $g(h)$. Observations indicate that a region with a length scale of 100 km or more is required to define $g(h)$ \cite{Thorndike:1975}, whereas the features that result from ridging and rafting extend over {\em up to} a few tens of meters in general \cite{VW08}. Hence, one could construct a theory that neglects the details of the collisions, but takes their net effect into account to study the geophysical-scale evolution of $g(h)$. This line of reasoning led us to use an analogy with Brownian motion to interpret $\psi$ \cite{TW2015}. Now we describe  this approach.

The classical problem of Brownian motion concerns the motion of a pollen grain in water \cite{Reif, chandra1943}. The collisions with the water molecules effect the motion of the pollen grain.  Given the length-scale separation between the pollen grain and the solvent molecules, there are an enormous number of solvent-grain collisions over the time scale of the evolution of the pollen grain. Hence, one does not take the individual collisions into account when describing its motion, but only their (appropriately averaged) net effect.

We view the short length and time scales of individual mechanical processes (ridging and rafting) relative to the overall evolution of $g(h,t)$\footnote{To be explicit in equation \ref{eqn:psi} we write the time dependence, but this is implicit elsewhere.} in direct analogy to the collisions of water molecules with a Brownian particle, and thus write $\psi$ as
\be
\psi(h,t) = \int_{0}^{\infty} \left[g(h^\prime,t)\, w(h,h^\prime) - g(h,t)\, w(h^\prime,h) \right] \, dh^\prime.
\label{eqn:psi}
\ee
Thus, we interpret the mechanical redistribution of ice thickness as the differential form of the Chapman-Kolmogorov equation, or a Master equation.  The transition probabilities per unit time $w(h,h^\prime)$ and $w(h^\prime,h)$  represent deformation processes changing ice from thickness $h^\prime$ to $h$ and from $h$ to $h^\prime$ respectively, and $w(h,h^\prime) = w(h^\prime,h)$. 
We Taylor expand equation \ref{eqn:psi} thereby transforming equation \ref{eqn:thorndike} to
\be
\frac{\partial g}{\partial t} = - \nabla \cdot ( \boldsymbol{u} g) - \frac{\partial}{\partial h} \left(f g\right) + k_1 \, \frac{\partial g}{\partial h} + k_2 \, \frac{\partial^2 g}{\partial h^2},
\label{eqn:newg(h)}
\ee
where
\begin{widetext}
\be
k_1 = \int_{0}^{\infty} \left|h^\prime - h\right| \, w(h,h^\prime) \, dh^\prime  \, \mathrm{~and~} \,
k_2 = \int_{0}^{\infty} \frac{1}{2}\left|h^\prime - h\right|^2 \, w(h,h^\prime) \, dh^\prime.
\ee
\end{widetext}
Equation \ref{eqn:newg(h)} is a Fokker-Planck-like equation 
that describes the evolution of the probability density $g(h,t)$.   Here, $k_1$ and $k_2$ represent the first and second moments of thickness transition events, which because of our core framework that the events that change the thickness occur very rapidly relative to the overall changes in $g(h,t)$, are constants.  

We nondimensionalize this equation by choosing $H_{eq}$ as the vertical length scale; $L$ as the horizontal length scale; $U_0$ as the velocity scale for the horizontal ice velocity; $t_m = L/U_0$ as the time scale for advection of ice floes; $t_D = H_{eq}^2/\kappa$, where $\kappa$ is the thermal diffusivity of ice, as the diffusion time scale; and $t_R \sim 1/\dot\gamma$, where $\dot \gamma$ is the collisional strain rate, as the relaxation time scale. Hence, the remaining terms have the following scalings: $f_0 = H_{eq}/t_D$, $\widetilde{k_{1}} = H_{eq}/t_R$, and $\widetilde{k_{2}} = H_{eq}^2/t_R$. Maintaining the pre-scaled notation and noting that $t_R \sim t_m$, equation \ref{eqn:newg(h)} is:
\be
\frac{\partial g}{\partial t} = - \nabla \cdot ( \boldsymbol{u} g) + \frac{\partial}{\partial h} \left[\left(k_1 - \tau f\right) g\right] + \frac{\partial^2}{\partial h^2}\left(k_2 g\right),
\label{eqn:fpt_scaled}
\ee
where $\tau \equiv t_m/t_D$. When $\boldsymbol{u} g(h)$ is solenoidal in the domain $R$ equation \ref{eqn:fpt_scaled} becomes
\be
\frac{\partial g}{\partial t} = \frac{\partial}{\partial h} \left(\phi g\right) + \frac{\partial^2}{\partial h^2}\left(k_2 g\right),
\label{eqn:fpt2}
\ee
where $\phi = k_1 - \tau f$. Equation \ref{eqn:fpt2} is a Fokker-Planck equation for $g(h,t)$. In this paper, we discuss the analytical and numerical solutions to equation \ref{eqn:fpt2} with a particular focus on the climatological evolution of the thickness distribution.  

\section{Analytical Solutions}
\subsection{Steady Solution}
A unique steady solution to equation \ref{eqn:fpt2} was obtained in \cite{TW2015} as follows.  The thermal growth rate was taken to be the solution to the ideal Stefan problem (see e.g., \cite{Worster:2000}) viz., $f = 1/S h$,  where $S$ is Stefan number defined as $S \equiv L_i/c_p \Delta T$ with $L_i$, $c_p$ and $\Delta T$ the latent heat of fusion of ice, specific heat of ice at constant pressure and the temperature difference across the ice layer, respectively. Using the boundary conditions $g(0) = g(\infty) = 0$ the steady state solution is 
\be
g(h) = {\cal N}(q) h^q e^{-h/H},
\label{eqn:fpt_ss_soln}
\ee
where $q = \tau/k_2 S = \epsilon/k_2$ and $H = k_2/k_1$. The prefactor, ${\cal N}(q) = \left[H^{1+q} \Gamma(1+q)\right]^{-1}$, is the normalization constant with $\Gamma(x)$ the Euler gamma function. Finally, we note that for a given $q$ and $H$, this solution is unique. 

Setting to zero the first derivative of  equation \ref{eqn:fpt_ss_soln} with respect to $h$ yields ${h = H_{eq}}$ as 
\be
H = \frac{1}{q} H_{eq},  
\label{eqn:heq}
\ee
thus also providing a derivation of Thorndike's dimensionless function (equation \ref{eqn:H-Heq}) from our steady state solution.

\subsection{Time-dependent Solution}

As a first step in understanding how the thickness distribution is driven by climatological forcing we introduce a simple model for the growth rate, with $f = 1$ and $-1$ during growth and melt seasons respectively. We emphasize that this model is intended to be pedagogical, as it does not accurately model the winter growth rate, which depends on the thickness. 
We use Chandrasekhar's method \cite{chandra1943} to first obtain the fundamental solution to equation \ref{eqn:fpt2}. The method involves computing the characteristics of the advective part of the equation, in which $\phi$ is now a constant, along which we write the resulting--diffusion--equation (see Appendix 1); 
\be
\frac{\partial g}{\partial s} = k_2 \frac{\partial^2 g}{\partial y^2},
\label{eqn:fpt4}
\ee
where $y =  h + \phi \, t$ and $s = t$, and the boundary conditions are $g(y = \phi \, s,s) = g(y = \infty,s) = 0$. The Green's function that satisfies these conditions is 
\be
\begin{split}
g(y,s;y_0) &= \frac{1}{\sqrt{4 \pi k_2 s}} \exp\left({-\frac{\left(y - y_0\right)^2}{4 k_2 s}}\right) \\
& - \frac{1}{\sqrt{4 \pi k_2 s}} \exp\left({-\frac{\left(y+ y_0\right)^2}{4 k_2 s} + \frac{\phi y_0}{k_2}}\right),
\end{split}
\label{eqn:FS}
\ee
and thus, the time-dependent solution for a given initial condition $g_0(y_0)$ is given by
\be
g(y,s) = \int_0^{\infty} g(y,s; y_0)\, g_0(y_0)\, dy_0.
\label{eqn:soln1}
\ee
For $g_0(y_0) = N y_0 e^{-by_0}$, the solution in terms of $h$ and $t$ is
\begin{widetext}
\be
\begin{split}
g(h,t) &= \frac{N}{2} (h+\phi t) e^{\left(-b (h+\phi t) + \frac{\theta^2 b^2}{4}\right)} \left[1 + \erf(\beta_1)\right] 
 - \frac{N}{2 \sqrt{\pi}} \theta e^{\left(-b (h+\phi t) + \frac{\theta^2 b^2}{4}\right)} \left\{-e^{-\beta_1^2} + \frac{\sqrt{\pi} \theta b}{2} \left[1 + \erf(\beta_1)\right]\right\} \\
& - \frac{N}{2 \sqrt{\pi}} \theta e^{\left(-\frac{\phi (h+\phi t)}{k_2} + b (h+\phi t) + \frac{\gamma^2}{4}\right)} \left[ e^{-\beta_2^2} + \frac{\sqrt{\pi} \gamma}{2} \erfc(\beta_2)\right] 
 + \frac{N}{2} (h+\phi t) e^{\left(-\frac{\phi (h+\phi t)}{k_2} + b (h+\phi t) + \frac{\gamma^2}{4}\right)}  \erfc(\beta_2),
\end{split}
\label{eqn:full_solution}
\ee
\end{widetext}
where $\theta = \sqrt{4 k_2 t}$, $\beta =  (h+\phi t)/\theta$, $\beta_1 = \beta - \alpha b/2$, $\beta_2 = \beta - \gamma/2$, $\gamma = \phi \theta/k_2 - b \theta$, and the error function (complimentary error function) is $\erf(X)$ ($\erfc(X)$).  Importantly, equation \ref{eqn:full_solution} demonstrates that even in the case of growth rate being independent of thickness, multiple time scales are generated due to the interaction between thermal and mechanical processes. 

\section{Numerical Solutions}
Seasonality and climate forcing is introduced by coupling equation \ref{eqn:fpt2} to the one-dimensional thermodynamic model of Eisenman \& Wettlaufer (EW09) \cite{EW09}.  We discretize equation \ref{eqn:fpt2} using the standard second-order finite-difference formulae and it is integrated in time using the semi-implicit Crank-Nicolson scheme. The equation for $f = f(h,t)$ is:
\be
\begin{split}
f &= \frac{1}{\rho_i \, L_i \, f_0} \left[-\left(1-\alpha\right) F_S + F_0 + \sigma_T T  - \Delta F_0 - F_B \right] \\
& - \frac{1}{f_0} \nu_0 h,
\label{eqn:ew09_1}
\end{split}
\ee
where $\rho_i$ is the density of ice, $F_S(t)$ is the incoming shortwave radiative flux, $F_0(t) = \sigma_0- F_{L}(t) - F_{SH}(t) - F_{LH}(t)$, $F_{L}(t)$ is the incoming longwave radiative flux, $F_{SH}(t)$, $F_{LH}(t)$ are the turbulent specific and latent heat fluxes at the upper surface, $\Delta F_0$ is the controlled flux perturbation at the upper surface (representing greenhouse gas forcing), and $F_B$ is the oceanic heat flux at the bottom surface. A linearized form of the Stefan-Boltzmann law is used for the outgoing longwave radiative flux and is given by $\sigma_0 + \sigma_T \,T(h,t)$, where $T(h,t)$ is the temperature of the upper surface. Ice export is $10\%$ per year and represented by $\nu_0 h$.  Whilst we have neglected the advection term in equation (\ref{eqn:fpt2}), we have incorporated the mean effect of advection on ice export in this manner, but it is not the same as incorporating the full ice velocity field.   However, ignoring export leads to relatively minor quantitative changes and no qualitative changes to the results presented here. Finally, we note that because $S$ is large for ice, the energy balance across $h$ is global and hence, for example,  $\Delta F_0 = 2$ Wm$^{-2}$, $F_B=0$ and $\Delta F_0 = 0$, $F_B=2$ Wm$^{-2}$ are equivalent. 

\subsection{Sea Ice Growth Rate}
\begin{figure}
\centering
\includegraphics[trim = 0 0 0 0, clip, width = 0.9\linewidth]{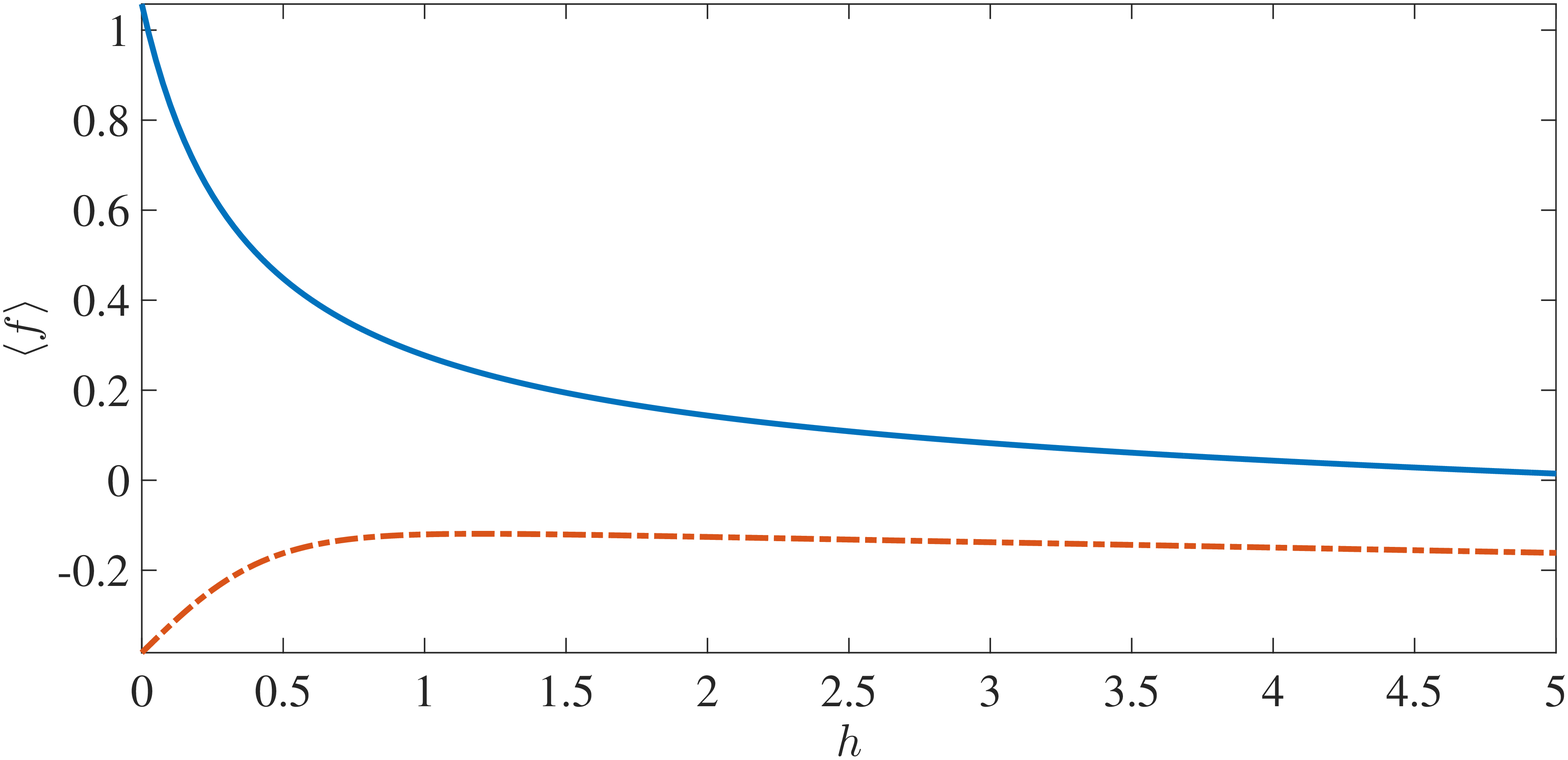}
\caption{Typical growth rate profiles for winter (solid line) and summer (dashed line). }
\label{fig:growth}
\end{figure}

The typical growth rates from the EW09 model for winter and summer are shown in figure \ref{fig:growth}.
Clearly, as this is a Stefan problem the growth rate decreases with increasing thickness, due to the fact that growth rate depends on the amount of heat conducted through the ice layer, which decreases with increasing thickness. The growth rates shown here are similar to those obtained from MU71 \cite{Thorndike:1975, MU71}. The melt rate is constant for all thicknesses except for $h < 1$, which can be attributed to the accelerated melting of thin ice because of the ice-albedo feedback.  The feedback is captured here through the $h$ dependence of the albedo, where the characteristic length scale is the inverse of the spectrally averaged Beer's extinction coefficient $\lambda$ = 0.67 m \cite{MU71}.   This becomes particularly important as the ice cover thins and $h \approx \lambda$ because our thermodynamic model does not account for the fraction of energy penetrating into the water column and effectively increasing $F_B$.  

\subsection{Evolution of the Mean Thickness}
\begin{figure}
\centering
\includegraphics[trim = 0 0 0 0, clip, width = 1\linewidth]{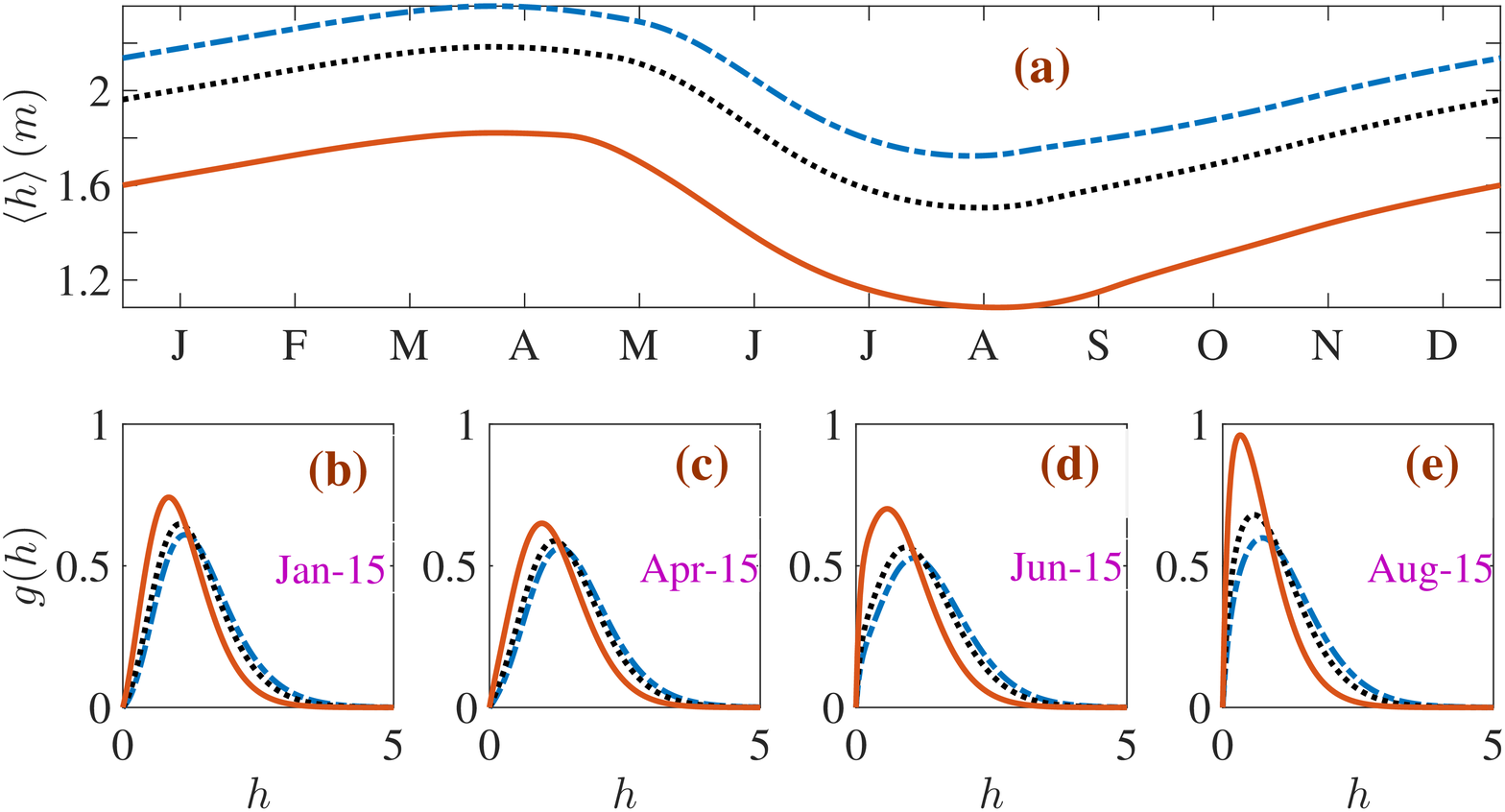}
\caption{Seasonal evolution of the mean thickness  $\left<h\right>$ and $g(h)$ versus $\Delta F_0$, and $H_{eq} = 1.5$ m throughout. Dash-dotted line: $\Delta F_0 = 2$ Wm$^{-2}$, ${\left<h\right>}_{\text{max}}$=2.36 m, ${\left<h\right>}_{\text{min}}$=1.72 m; dotted line: $\Delta F_0 = 15$ Wm$^{-2}$, ${\left<h\right>}_{\text{max}}$=2.18 m, ${\left<h\right>}_{\text{min}}$=1.50 m; and solid line: $\Delta F_0 = 50$ Wm$^{-2}$, ${\left<h\right>}_{\text{max}}$=1.82 m, ${\left<h\right>}_{\text{min}}$=1.08m. }
\label{fig:g_phase1}
\end{figure}

The mean thickness is defined as:
\be
\langle h(t)\rangle =  \int_0^{\infty} h \,g(h,t) \, dh,
\label{eqn:mean_thickness}
\ee
in the same manner as the mean values $\langle X \rangle$ of all quantities $X$. Figure \ref{fig:g_phase1}(a) shows the seasonal behavior of the dimensional $\left<h\right>$.
For all values of $\Delta F_0$ the seasonal cycle of $\left<h\right>$ is qualitatively the same, with the maximum, ${\left<h\right>}_{\text{max}}$, at the end of the growth season in early April, and the minimum, ${\left<h\right>}_{\text{min}}$, at the end of the melt season in August.   This behavior is in general agreement with observations and with solely thermodynamic models \cite{MU71, EW09}.   Importantly, however, for greenhouse gas forcing roughly twice that at which the thermodynamic only component of this model transitions from the seasonal ice state to the ice free state (see Fig. 3 of \cite{EW09}), here we are still in the perennial state.  This exhibits the crucial role that ice mechanics plays in maintaining the ice cover by redistributing thin ice to thick ice--far more rapidly than can thermal growth alone.  Indeed, it is not until  $\Delta F_0$ reaches approximately six times the thermodynamic only transition to the ice free state that the annual mean $\left<\overline{h}\right> \approx \lambda$ (see Fig. \ref{fig:h_fb}).   

\subsection{Seasonal and Climatological Changes in $g(h)$}
\begin{figure}
\centering
\includegraphics[trim = 0 0 0 0, clip, width = 0.9\linewidth]{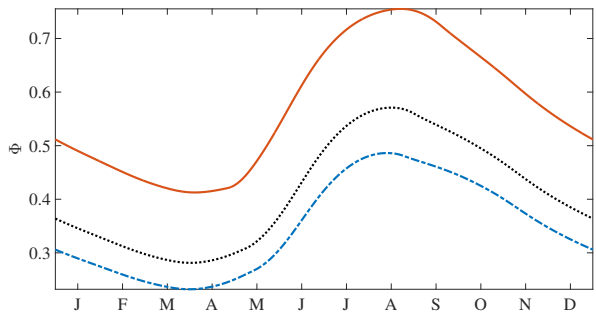}
\caption{Seasonal evolution of the thin-ice fraction with increasing greenhouse gas forcing. Dash-dotted line: $\Delta F_0 = 2$ Wm$^{-2}$; dotted line: $\Delta F_0 = 15$ Wm$^{-2}$; and solid line: $\Delta F_0 = 50$ Wm$^{-2}$.}
\label{fig:phi_compare}
\end{figure}

For the purpose of discussing the changes in $g(h)$, we define `thin' ice as ice of thickness $h \le 1$ and `thick' ice as ice of thickness $h>1$. The fraction of thin ice is
\be
\Phi = \int_0^1 g(h)\, dh.  
\ee
Whereas $\Phi$ increases by about a factor of two from the winter minimum to the summer maximum (figure \ref{fig:phi_compare}), roughly independent of $\Delta F_0$, the minimum and the maximum also increase by roughly a factor of two as $\Delta F_0$ increases.  

Figures \ref{fig:g_phase1}(b) -- (e) show $g(h)$ at the middle and end of the growth and melt seasons respectively.   For all $\Delta F_0$, as winter progresses and $\Phi$ decreases, both thermal growth and mechanical redistribution drive the spread and rightward motion of $g(h)$ to create more thick ice.  As $\Phi$ increases during the melt season, 
$g(h)$ contracts towards thinner ice and the skewness increases, both of which are enhanced substantially as $\Delta F_0$ increases.  

We make a qualitative comparison of $g(h)$ with the recent satellite observations from CryoSat-2 \cite{kwok2015} in figure \ref{fig:kwok}. The data has been averaged over the periods shown in figure \ref{fig:kwok}(a), whereas data from the model in \ref{fig:kwok}(b) are for the particular days shown. The observed spreading of $g(h)$ during winter is explained as above within the framework of the theory; ice growth makes thicker the ice that is formed and subsequently deformed, shifting the peak to the right and broadening the distribution, a behavior that is suppressed as $\Delta F_0$ increases.  It is the continual deformation of thinner ice to make thicker ice that maintains a thicker ice pack than would be predicted by thermodynamic only models.  

\begin{figure}
\centering
\includegraphics[trim = 0 0 0 0, clip, width = 1\linewidth]{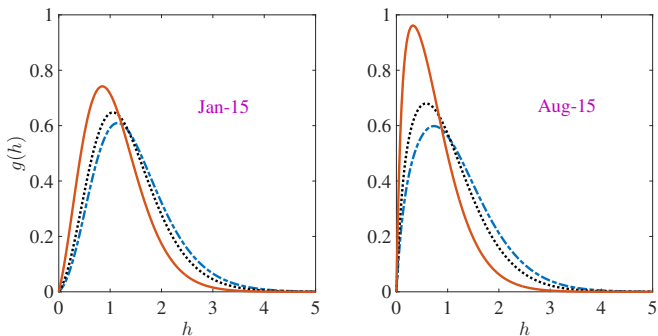}
\caption{Seasonal evolution of $g(h)$ with increasing greenhouse gas forcing. Dash-dotted line: $\Delta F_0 = 2$ Wm$^{-2}$; dotted line: $\Delta F_0 = 15$ Wm$^{-2}$; and solid line: $\Delta F_0 = 50$ Wm$^{-2}$.}
\label{fig:g_phase2}
\end{figure}

\begin{figure}
\centering
\includegraphics[trim = 0 0 0 0, clip, width = 1\linewidth]{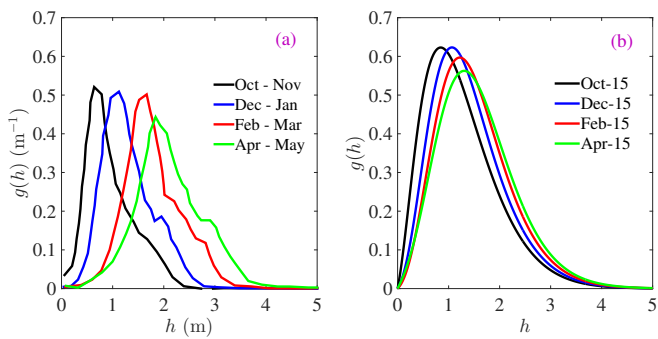}
\caption{Qualitative comparison with CryoSat-2 observations \cite{kwok2015}. (a) $g(h)$ for first-year ice for the year 2010-2011; the data were obtained over the periods indicated. (b) $g(h)$ for the whole thickness range for $\Delta F_0 = 2$ Wm$^{-2}$ from our model on the particular days as indicated.}
\label{fig:kwok}
\end{figure}

\subsection{Albedo}
\begin{figure}
\centering
\includegraphics[trim = 0 0 0 0, clip, width = 0.85\linewidth]{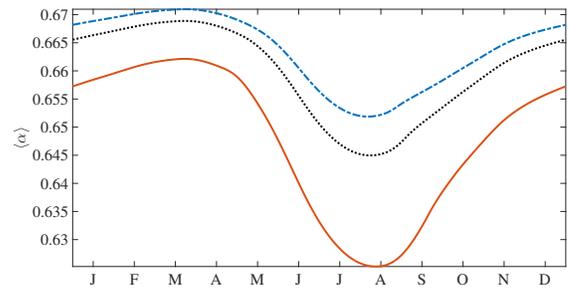}
\caption{Seasonal evolution of the mean albedo with increasing greenhouse gas forcing. Dash-dotted line: $\Delta F_0 = 2$ Wm$^{-2}$; dotted line: $\Delta F_0 = 15$ Wm$^{-2}$; and solid line: $\Delta F_0 = 50$ Wm$^{-2}$. }
\label{fig:albedo}
\end{figure}

Importantly, once $g(h)$ is known, all thickness dependent moments can be calculated.
A quantity of keen interest is the albedo, whose summer mean values are difficult to model because of the concurrent presence of a wide range of ice thicknesses in the basin. 
We plot the seasonal evolution of the mean albedo $\left<\alpha\right>$ as a function of $\Delta F_0$ in figure \ref{fig:albedo}.
For example, when $\Delta F_0 = 2$ Wm$^{-2}$ we see that $\left<\alpha\right>$ reaches a maximum ($0.671$)  at the end of the growth season, and a minimum ($0.652$) at the end of the melt season; a seasonal difference in the extreme values of only $2.9\%$, but this translates into a large variation in surface heat balance \cite{EUW:2007}.   Figures \ref{fig:phi_compare} and \ref{fig:albedo} show that $\Phi$ and $\left<\alpha\right>$ are anticorrelated; and a close observation of the plots reveals a phase difference between them with $\left<\alpha\right>$ leading.  Importantly, as $\Delta F_0$ increases so too does the amplitude of the seasonal cycle, the peak to peak variation of which has a substantial impact on the radiative forcing and hence the ice thickness. 

\subsection{Effects of the surface radiative flux forcing}
\begin{figure}
\centering
\includegraphics[trim = 0 0 0 0, clip, width = 0.9\linewidth]{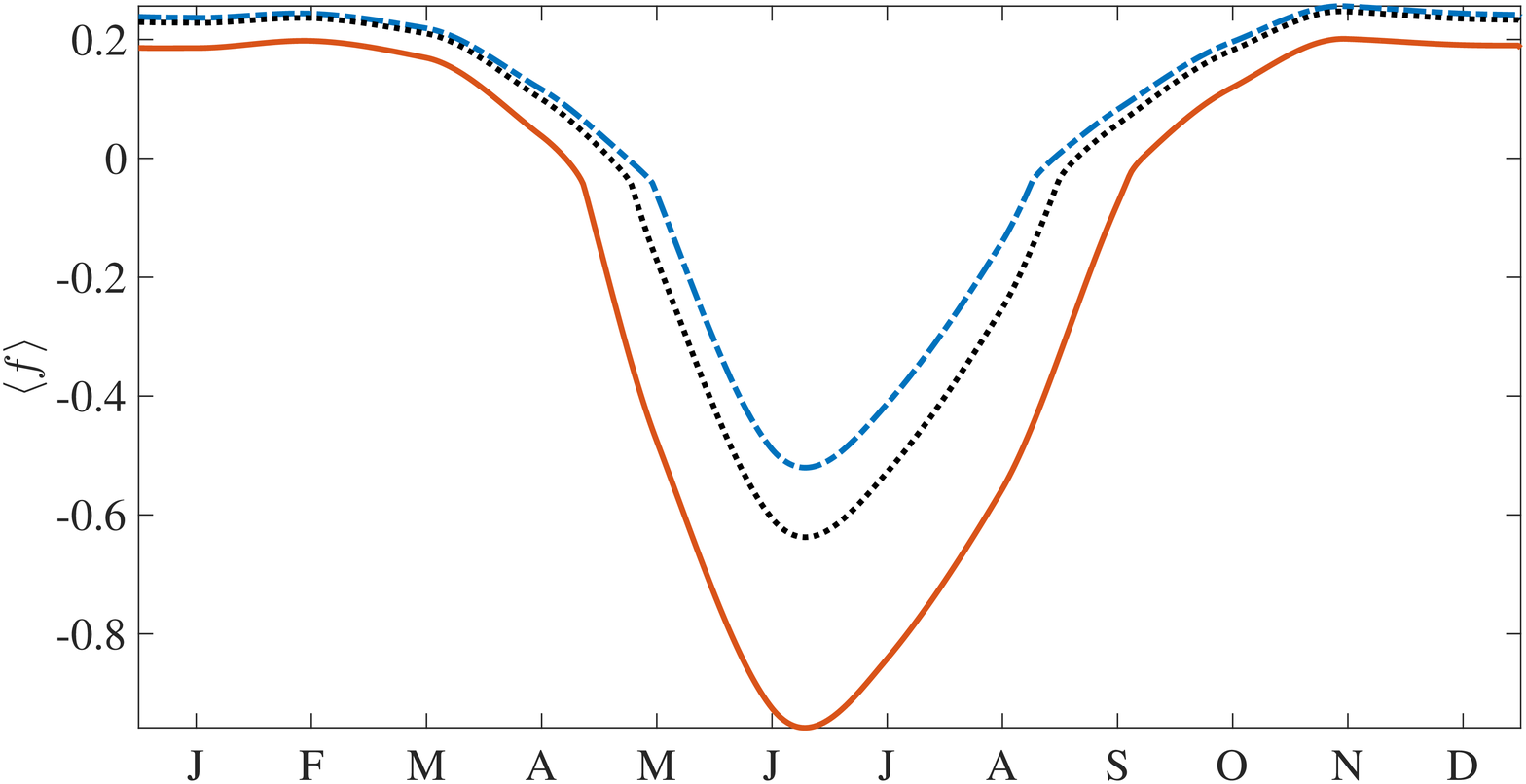}
\caption{Seasonal changes in the mean growth rate of ice with increasing greenhouse gas forcing. Dash-dotted line: $\Delta F_0 = 2$ Wm$^{-2}$; dotted line: $\Delta F_0 = 15$ Wm$^{-2}$; and solid line: $\Delta F_0 = 50$ Wm$^{-2}$.}
\label{fig:mean_growth}
\end{figure}

\begin{figure}
\centering
\includegraphics[trim = 0 0 0 0, clip, width = 0.9\linewidth]{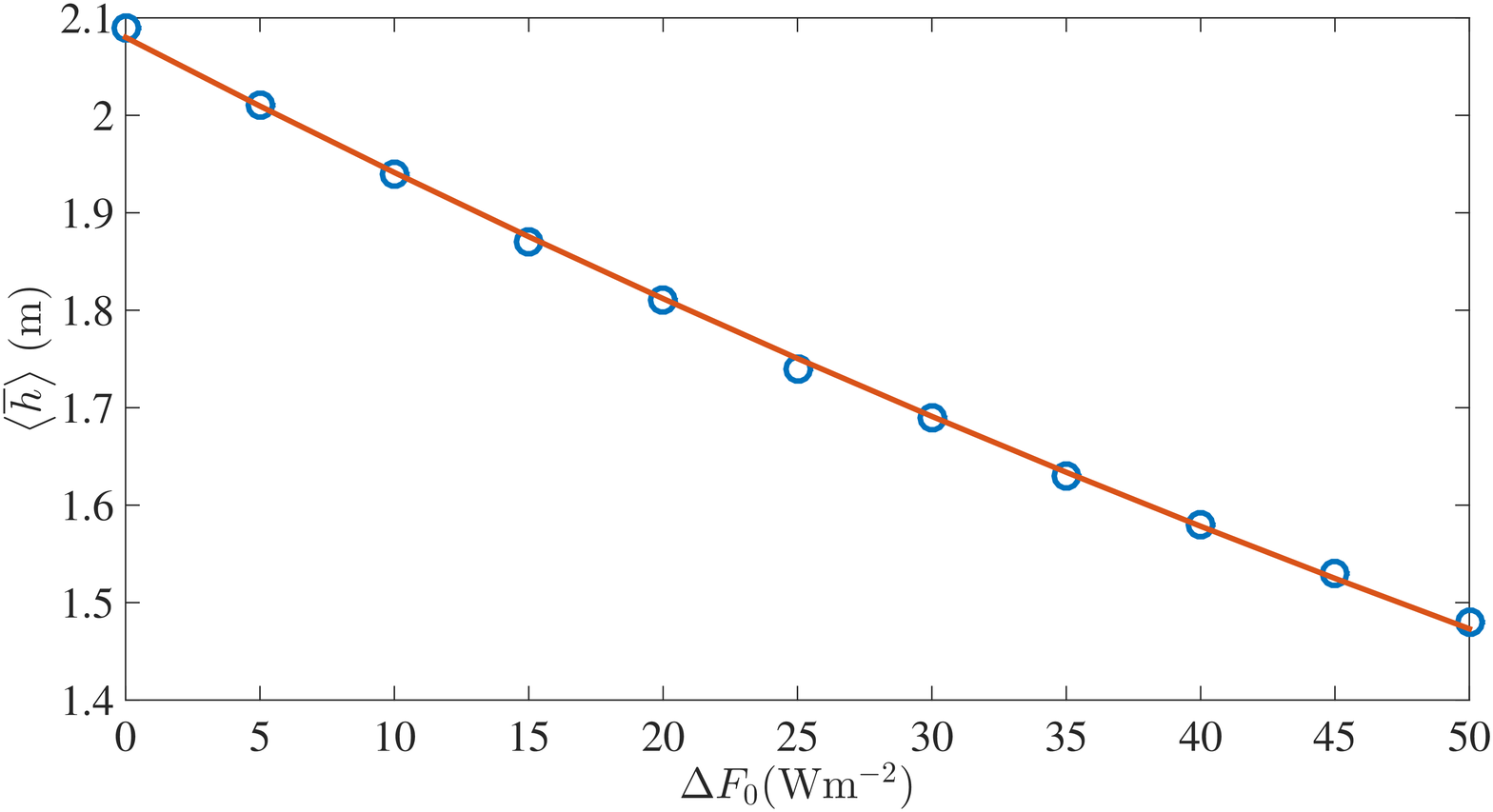}
\caption{Effect of increasing $\Delta F_0$ on $\left<\overline{h}\right>$. Circles: Simulation; solid line: $\left<\overline{h}\right> = 2.08 \times \exp{\left(-0.0069 \times \Delta F_0 \right)}$. The mean thickness becomes $0.66$ m $\approx \lambda$ when $\Delta F_0 = 164$ Wm$^{-2}$.}
\label{fig:h_fb}
\end{figure}

The effect of $\Delta F_0$ on $g(h)$ can be understood by considering equation \ref{eqn:ew09_1}. An increase in $\Delta F_0$ results in a smaller growth rate during winter and a higher melt rate during summer, leading to an increase in $\Phi$ for all seasons (figure~\ref{fig:phi_compare}). Figure \ref{fig:mean_growth} shows that the mean growth rate shifts downward with increasing $\Delta F_0$, but the curves are not phase shifted. 
This shift in $\left<f\right>$ is associated with the increase in $\Phi$ for all seasons.  Figure \ref{fig:mean_temp} shows the $\Delta F_0$ dependence of the seasonal cycle of the mean ice surface temperature $\left<T\right>$.  It is seen that as $\Delta F_0$ increases, so too does $\left<T\right>$ and thus the winter growth rate decreases.  Moreover, the time period during which the upper surface ablates increases with $\Delta F_0$. This combination of effects leads to a decrease in $\left<\overline{h}\right>$. Figure \ref{fig:h_fb} shows that $\left<\overline{h}\right>$ decreases exponentially with increasing $\Delta F_0$ over the range simulated, and thus should vanish monotonically in the absence of some other feedback, in qualitative, but as noted above not quantitative, agreement with solely thermodynamic models \cite{MU71, EW09}.  

Finally, given the important shift in the transitions of the ice states (perennial to seasonal to ice free) from solely thermodynamic models to the full mechanical and thermodynamic treatment of this theory, the 
response of the ice pack to a radiative flux perturbation is clearly different between these approaches.  We have quantified the relaxation time scales for different initial conditions (see Appendix 3) and find that there is a range of thin ice fractions, $\Phi \approx 0.3 - 0.6$, for which the relaxation time scale of the ice pack is approximately 50\% that of thermodynamic only models; viz., $\sim$ 4 years rather than $\sim$ 10 years.   For distributions with much more thick ice the response time scales are controlled by mechanical deformation of thick ice and hence can be much longer. 

\section{Conclusion}
Using concepts and methods from statistical physics we have transformed the theory of the sea ice thickness distribution, $g(h)$, of Thorndike \emph{et al.} \cite{Thorndike:1975} into a solvable Fokker-Planck-like equation.  We have solved the new equation both analytically and numerically using different models for the thermodynamic growth rate $f$ to understand the climatological evolution of $g(h)$. In the simplest case, $f = \pm 1$ for the growth and melt seasons, and this yields an analytical solution (equation \ref{eqn:full_solution}). The solution shows that the interaction of thermal and mechanical processes during the evolution of $g(h)$ leads to the generation of multiple time scales, which in turn affect the evolution. Thus, as previously suggested by Thorndike \cite{Thorndike:1992}, we do in fact find that $g(h)$ and its moments relax on different time scales, which has important geophysical consequences. 

A climatological suite of calculations was performed by coupling the Fokker-Planck equation to the thermodynamic model of Eisenman and Wettlaufer \cite{EW09}.  
The temporal and time averaged $g(h)$ from our model are in good agreement with the recent satellite measurements over the Arctic basin \cite{kwok2015, TW2015}.  
As in solely thermodynamic models \cite{MU71, EW09}, we find that the stationary state has a mean thickness, $\left<h\right>$, reaching a maximum in early April, which is the end of the growth season, and a minimum in early August, which is the end of the melt season.  Due to the combined effects of thermodynamics and mechanics, $g(h)$ spreads during the growth season and contracts during the melt season. As greenhouse gas forcing, $\Delta F_0$, increases this contraction is enhanced, with a larger skewness and a sharper peak at lower thicknesses.  However, this model remains in the perennial ice state
for $\Delta F_0$ approximately twice that at which its thermodynamic component transitions from the seasonal ice state to the ice free state.  
This exhibits the crucial role that ice mechanics plays in maintaining the ice cover by redistributing thin ice to thick ice; intuitively, doubling the thickness of thin ice by ridging occurs instantaneously \cite{VW08} relative to doubling it by thermal growth.  Clearly, by such a stage the ice-covered fraction of the Arctic Ocean may be vastly smaller than at present.  Nonetheless, these physical processes will persist until other effects, such as changing boundary conditions at lower latitudes, take over.  For example, although it is not until  $\Delta F_0$ reaches approximately six times the thermodynamic only transition to the ice free state that the exponential decay of the annual mean, $\left<\overline{h}\right>$, is reduced to the decay scale of shortwave radiation $\lambda \approx 0.66$ m (Fig. \ref{fig:h_fb}).   

The seasonal behavior of the thin-ice fraction $\Phi$ is anticorrelated with the behavior of $\left<h\right>$, which is correlated with the evolution of the mean albedo $\left<\alpha\right>$. 
The surface radiative flux perturbation $\Delta F_0$  impacts $g(h)$ by decreasing the mean growth rate and the seasonally averaged mean thickness $\left<\overline{h}\right>$, thereby 
leading to an increase in $\Phi$. Depending on the initial $\Phi$, the relaxation times for $g(h)$ to reach a stationary state starting from an arbitrary initial condition range from $\sim$ 4 - 10 years.  Importantly, for a range of thin ice fractions, $\Phi \approx 0.3 - 0.5$, there is a minimum in the relaxation time of $\sim$ 4 years, which is approximately 50\% that of thermodynamic only models.  For distributions with much more thick ice the response time scales are controlled by mechanical deformation of thick ice and thus become much longer. 

The results presented here demonstrate veracity of using the methods and concepts of statistical mechanics to study the geophysical-scale evolution of Arctic sea ice.  
As described in the introduction, the CMIP5 models poorly represent the spatial patterns of ice thickness \cite{Stroeve:2014}.  The concept of the original theory of $g(h)$ due to Thorndike \emph{et al.} \cite{Thorndike:1975} was to avoid the complexities of unknown ice rheologies in the equations of motion for the ice cover, and to produce a climatologically relevant and easily implementable probability density function of this core geophysical scale variable for the polar ocean.  However,  implementation was difficult because of the intransigence of the redistribution function $\psi$. Having solved this problem in our theory, we find solutions that are in good agreement with satellite observations. Therefore, using the present treatment for $g(h)$ in climate models should lead to a more realistic representation of Arctic sea ice within them.  The thermodynamic component used here \cite{EW09} reproduces the seasonal cycle of Maykut \& Untersteiner \cite{MU71}, which is the starting point for all subsequent simplifications of the thermodynamics used in climate models.  Therefore, the implementation of our approach, which captures both the mechanics and the thermodynamics, in comprehensive models should be of interest.  

\begin{acknowledgements}
ST acknowledges a NASA Graduate Research Fellowship.  JSW acknowledges NASA Grant NNH13ZDA001N-CRYO, Swedish Research Council grant no. 638-2013-9243, and a Royal Society Wolfson Research Merit Award for support.  As Yalies working in statistical mechanics, we took inspiration from Leo Kadanoff's deep understanding of the field \cite{Kadanoff:2014}, and his constant encouragement to explore its vast tendrils; we hope he would have enjoyed this effort.

\end{acknowledgements}

\section*{Appendix 1: Exact solution of the Fokker-Planck equation} \label{app:exact_solution}
The Fokker-Planck-like equation for the thickness distribution, $g(h)$, for constant thermal growth rate is
\be
\frac{\partial g}{\partial t} = \phi \frac{\partial g}{\partial h} + k_2 \frac{\partial^2 g}{\partial h^2},
\label{eqn:original}
\ee
where $\phi = k_1 - \tau f$, with the boundary conditions $g(h=0,t) = g(h=\infty,t)=0$. The characteristics of the advective part of equation \ref{eqn:original} are
\be
\frac{dh}{dt} = - \phi; \, \frac{ds}{dt} = 1.
\ee
Hence, using the transformation $y = h + \phi t$ and $s = t$ in equation \ref{eqn:original} gives
\be
\frac{\partial g}{\partial s} = k_2 \frac{\partial^2 g}{\partial y^2},
\label{eqn:heat}
\ee
with the transformed boundary conditions; $g(y = \phi s, s) = g(y = \infty, s) = 0$. Equation \ref{eqn:heat} is the diffusion equation for $g$ along the characteristics. 

We solve equation \ref{eqn:heat} by first finding its Green's function, $g(y,s; y_0)$, which satisfies the following expression\
\be
\frac{\partial g}{\partial s} - k_2 \frac{\partial^2 g}{\partial y^2} = \delta\left(y-y_0\right) \, \delta(s),
\label{eqn:greens}
\ee
where, $\delta(X)$ is the Dirac-delta function. Following Duffy \cite{duffy2015}, one can seek the Green's function in the form
\be
g(y,s; y_0) = G(y,s; y_0) + u(y,s),
\ee
where $G(y,s;y_0)$ is the free-space Green's function given by
\be
G(y,s;y_0) = \frac{1}{\sqrt{4 \pi k_2 s}} \exp\left[-\frac{(y-y_0)^2}{4 k_2 s}\right],
\ee
and $u(y,s)$ is a homogeneous function that satisfies the following boundary condition at $y = \phi s$; 
\be
u(y = \phi s, s) = -G(y = \phi s, s; y_0),
\ee
which ensures the boundary condition $g(y = \phi s,s) = 0$ is enforced at all $s$. The function $u(y,s)$ satisfies equation \ref{eqn:heat} and can be shown to be  \cite{duffy2015}
\be
u(y,t) = -\frac{1}{\sqrt{4 \pi k_2 s}} \exp\left[-\frac{(y+y_0)^2}{4 k_2 s} + \frac{\phi y_0}{k_2}\right].
\ee
Once $g(y,s;y_0)$ is known, the solution corresponding to any initial condition $g_0(y_0)$ can be calculated as
\be
g(y,s) = \int_0^{\infty} g(y,s;y_0) \, g_0(y_0) \, dy_0.
\ee
For $g_0(y_0) = N y_0 e^{-by_0}$, the solution in terms of the original variables $h$ and $t$ is:
\be
\begin{split}
g(h,t) &= \frac{N}{2} (h+\phi t) e^{\left(-b (h+\phi t) + \frac{\theta^2 b^2}{4}\right)} \left[1 + \erf(\beta_1)\right] \\
 &- \frac{N}{2 \sqrt{\pi}} \theta e^{\left(-b (h+\phi t) + \frac{\theta^2 b^2}{4}\right)} \left\{-e^{-\beta_1^2} + \frac{\sqrt{\pi} \theta b}{2} \left[1 + \erf(\beta_1)\right]\right\} \\
& - \frac{N}{2 \sqrt{\pi}} \theta e^{\left(-\frac{\phi (h+\phi t)}{k_2} + b (h+\phi t) + \frac{\gamma^2}{4}\right)} \left[ e^{-\beta_2^2} + \frac{\sqrt{\pi} \gamma}{2} \erfc(\beta_2)\right] \\
 &+ \frac{N}{2} (h+\phi t) e^{\left(-\frac{\phi (h+\phi t)}{k_2} + b (h+\phi t) + \frac{\gamma^2}{4}\right)}  \erfc(\beta_2),
\end{split}
\ee
where $\theta = \sqrt{4 k_2 t}$, $\beta =  (h+\phi t)/\theta$, $\beta_1 = \beta - \theta b/2$, $\beta_2 = \beta - \gamma/2$, and $\gamma = \phi \theta/k_2 - b \theta$.  The error function and complimentary error function are $\erf(X)$ and $\erfc(X)$ respectively.

\section*{Appendix 2: Numerical Scheme}
The Fokker-Planck equation for the sea ice thickness distribution is
\be
\frac{\partial g}{\partial t} = \frac{\partial}{\partial h} \left(\phi g\right) + \frac{\partial^2}{\partial h^2}\left(k_2 g\right).
\label{eqn:fpt2_app}
\ee
To solve equation \ref{eqn:fpt2_app} numerically, we discretize it using the standard second-order finite difference formulae \cite{leveque2007}, and integrate it in time using the semi-implicit Crank-Nicolson method. This particular method is chosen for its stability and accuracy \cite{palleschi1990}. The finite difference form of equation \ref{eqn:fpt2_app} is
\begin{widetext}
\be
\frac{g_i^{n+1} - g_i^n}{\Delta t} = \frac{1}{2} \left[\frac{\phi_{i+1}^{n+1} g_{i+1}^{n+1} - \phi_{i-1}^{n+1} g_{i-1}^{n+1}}{2 \Delta h}\right]  
+ \frac{k_2}{2} \left[\frac{g_{i+1}^{n+1} - 2 g_i^{n+1} + g_{i-1}^{n+1}}{(\Delta h)^2}\right] 
 + \frac{1}{2} \left[\frac{\phi_{i+1}^{n} g_{i+1}^{n} - \phi_{i-1}^{n} g_{i-1}^{n}}{2 \Delta h}\right] 
 + \frac{k_2}{2} \left[\frac{g_{i+1}^{n} - 2 g_i^{n} + g_{i-1}^{n}}{(\Delta h)^2}\right],
\label{eqn:fpt_cn1}
\ee
\end{widetext}
where $\Delta h$ and $\Delta t$ are the ice thickness and time steps respectively. Here, $g_i^n$ corresponds to $g(h_i = i \Delta h, t_n = n \Delta t)$ where $h_i$ ($ i = 1, 2, 3, \ldots$) and $t_n$ ($n = 1, 2, 3, \ldots$) are the discrete values of $h$ and $t$, and similarly for the remaining terms. We rearrange equation \ref{eqn:fpt_cn1} as 
\begin{widetext}
\be
-\left(c_1 \phi_{i+1}^{n+1} + c_2\right) g_{i+1}^{n+1} + \left(1+2c_2\right) g_{i}^{n+1} + \left(c_1 \phi_{i-1}^{n+1}-c_2\right) g_{i-1}^{n+1} = 
\left(c_2+c_1 \phi_{i+1}^{n} \right) g_{i+1}^{n}
+  \left(1-2c_2\right) g_{i}^{n} + \left(c_2 - c_1 \phi_{i-1}^{n}\right) g_{i-1}^{n},
\label{eqn:fpt_cn_tridiag}
\ee
\end{widetext}
where $c_1 = \Delta t/4 \Delta h$ and $c_2 = k_2 \Delta t/ 2\left(\Delta h\right)^2$. Equation \ref{eqn:fpt_cn_tridiag} represents a tridiagonal system, which can be solved efficiently \cite{leveque2007}.

\begin{figure}[h]
\centering
\includegraphics[trim = 0 0 0 0, clip, width = 1\linewidth]{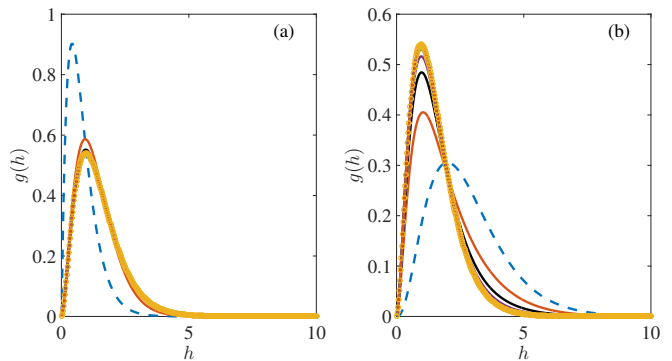}
\caption{Validation of the method. Panels (a) and (b) show that starting from different initial conditions (dashed lines) the solution converges to the unique steady state (circles). Here, $\Delta t = 0.01$, $\Delta h = 0.025$ and the total integration time $T_p = 400$ for the simulations.}
\label{fig:validation1}
\end{figure}

To validate the code, we solve equation \ref{eqn:fpt2_app} with the growth rate from the ideal Stefan problem \cite{Worster:2000}. The solution in this case is \cite{TW2015}
\be
g(h) = {\cal N}(q) h^q e^{-h/H}.
\ee
Thus, the test used is that starting from different initial conditions, this unique steady state solution should be reached.
We initialize $g(h)$ using: (a) $q = 1.05$, $H = 0.4$ and (b) $q = 2.5$, $H=0.8$. However, the values $q_{ss} = 1.84$ and $H_{ss} = 0.52$ are chosen to represent the final steady state solution, and hence values of $k_1$  and $k_2$ corresponding to $q_{ss}$ and $H_{ss}$ are used for the integration of equation \ref{eqn:fpt2_app}. For the test cases, we choose $\Delta t = 0.01$, $\Delta h = 0.025$, $T_p = 400$ as the total integration time, with $h_{min} = 0.01$ the smallest and $h_{max} = 10$ the largest discrete thicknesses. The boundary conditions imposed are; $g(h_{min}) = g(h_{max}) = 0$. Figure \ref{fig:validation1} shows that starting with the different initial conditions, the unique steady-state solution is reached.

\section*{Appendix 3: Seasonality in Growth Rate}
We introduce seasonality by solving the equation for the thermodynamic growth rate from the one-dimensional thermodynamic model of Eisenman \& Wettlaufer \cite{EW09} that is coupled to climatology.  Here, in order to make this paper reasonably self contained we summarize this energy balance model, in which the {\em dimensional form} \footnote{To avoid introduction of yet more notation here we simply state the dimensional form of the equation without specific redefining variables.} of the growth rate is
\be
\begin{split}
f(h,t) &= \frac{1}{\rho \, L_i}\left\{-\left[1-\alpha(h)\right] F_S(t) +F_0(t) + \sigma_T T(h,t)\right\} \\
&-\frac{1}{\rho \, L_i} \left[\Delta F_0 + F_B\right] - \nu_0 h, 
\label{eqn:ew09_1_app}
\end{split}
\ee
where $\rho_i$ is the density of ice, $F_S(t)$ is the incoming shortwave radiative flux, $F_0(t) = \sigma_0- F_{L}(t) - F_{SH}(t) - F_{LH}(t)$, $F_{L}(t)$ is the incoming longwave radiative flux, $F_{SH}(t)$ and $F_{LH}(t)$ are the turbulent specific and latent heat fluxes at the top surface, and $F_B$ is the oceanic heat flux at the bottom surface. A linearized form of the Stefan-Boltzmann law is used for the outgoing longwave radiation flux, and is given by $\sigma_0 + \sigma_T \,T(h,t)$.  Ice export is $10\%$ per year and represented by $\nu_0 h$.  Here, $T(h,t)$ is the upper surface temperature obtained from the flux balance 
\be
T(h,t) = -{\cal R}\left\{\frac{\left[1-\alpha(h)\right] F_s(t) - F_0(t) + \Delta F_0}{-k_i/h - \sigma_T}\right\},
\label{eqn:ew09_2}
\ee
where $k_i$ is the thermal conductivity of ice, and ${\cal R}(x)$ is the ramp function defined as
\[ {\cal R}(x) =
  \begin{cases}
    x       & \quad \text{if } x > 0,\\
    0  & \quad \text{if } x \le 0.\\
  \end{cases}
\]
The dependence of albedo on thickness is modelled using:
\be
\alpha(h) = \frac{\alpha_w + \alpha_i}{2} + \frac{\alpha_w-\alpha_i}{2} \tanh\left(-\frac{h}{\lambda}\right),
\label{eqn:ew09_3}
\ee
where $\alpha_i$ and $\alpha_w$ are the values of albedo for thickest ice and open water, respectively and $\lambda$ is inverse of the spectrally averaged extinction coefficient for Beer's law \cite{EW09}. The radiation climatology used to determine the values of $F_S(t)$, $F_{L}(t)$, $F_{SH}(t)$, and $F_{LH}(t)$ are from Maykut \& Untersteiner \cite{MU71}, as is $\lambda$. The values of $f(h,t)$ obtained are nondimensionalized by $f_0$ and then used when solving equation (11) of the main document or equation (\ref{eqn:fpt2_app2}) below. 

Choosing $H_{eq} = 1.5$ m, $L = 10^5$ m, $U = 0.1$ ms$^{-1}$ and $\kappa = 6.02 \times 10^{-7}$ m$^2$s$^{-1}$ gives $\tau = 0.27$, which is the value used throughout this study. The advection time scale corresponds to $t_m \approx$ 12 days. The number of days in a year is taken to be 360, which in non-dimensional units corresponds to $t=30$.

The values used for the constants are: $\rho_i = 917$ kgm$^{-3}$, $L_i = 333.4 \times 10^3$ Jkg$^{-1}$, $k_i = 2.2$ Wm$^{-1}$K$^{-1}$, $\sigma_0 = 316$ Wm$^{-2}$, $\sigma_T = 3.9$ Wm$^{-2}$K$^{-1}$, $\alpha_i = 0.68$ and $\alpha_w = 0.20$. The values of $k_1$ and $k_2$ corresponding to $q_{ss}$ and $H_{ss}$ are used throughout this study.

Convergence of the code for $\Delta F_0 = 2$ and $15$ Wm$^{-2}$ was ascertained from simulations with (i) $\Delta t = 0.05$, $0.01$ and $0.005$, for a fixed $\Delta h = 0.05$, and (ii) $\Delta h = 0.05$ and $0.025$ for a fixed $\Delta t = 0.01$. For each $\Delta F_0$, the convergence results obtained were identical in all cases. Hence, $\Delta t = 0.05$ and $\Delta h = 0.025$ were chosen for all the simulations.

\subsubsection*{Sensitivity of Surface Temperature to $\Delta F_0$}

\begin{figure}[h]
\centering
\includegraphics[trim = 0 0 0 0, clip, width = 1\linewidth]{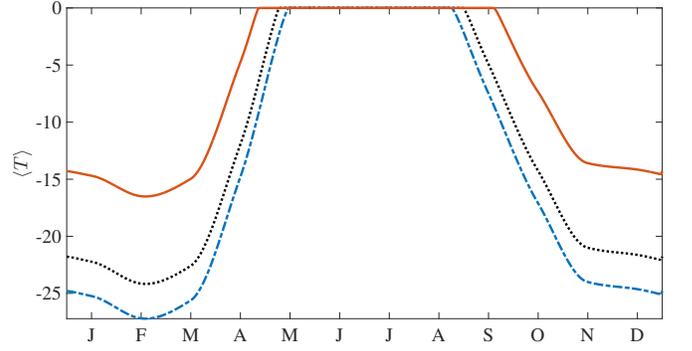}
\caption{Variation in the mean temperature with seasons for different $\Delta F_0$. Dash-dotted line: $\Delta F_0 = 2$ Wm$^{-2}$; dotted line: $\Delta F_0 = 15$ Wm$^{-2}$; and solid line: $\Delta F_0 = 50$ Wm$^{-2}$.}
\label{fig:mean_temp}
\end{figure}

Figure \ref{fig:mean_temp} shows the $\Delta F_0$ dependence of the seasonal cycle of the mean ice surface temperature $\left<T\right>$.  It is seen that as $\Delta F_0$ increases, so too does $\left<T\right>$ and thus the winter growth rate decreases.  Moreover, the time period during which the upper surface ablates increases with $\Delta F_0$.  This combination of effects leads to a decrease in the annually averaged mean thickness $\left<\overline{h}\right>$.

\subsubsection*{System Relaxation Time Scales}

We define the system relaxation time, $\Lambda_R$, as the time taken for $g(h,t)$ to evolve to a stationary state starting from an arbitrary initial condition. A knowledge of $\Lambda_R$ is
 important in answering the following question: Given an initial state of the ice pack, if there is a flux perturbation associated with a change in the environment, how quickly does the system forget its initial condition and reach a new stationary state? The variation of $\Lambda_R$ as a function of the initial condition may also help us understand the interaction between thermodynamics and mechanics that drives the system to the new stationary state. One possible way to answer this question would be to compute the relaxation times of the moments, but from Thorndike's \cite{Thorndike:1992} and our calculations 
(equation 17 of the main document) it is clear that even for constant thermodynamic growth rates, $g(h)$ and its moments relax on different time scales.

\begin{figure}[h]
\centering
\includegraphics[trim = 0 0 0 0, clip, width = 1\linewidth]{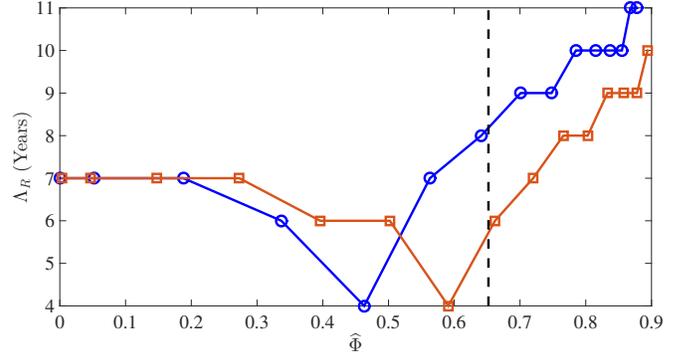}
\caption{Relaxation time $\Lambda_R$ as a function of thick-ice fraction $\widehat{\Phi}$. Circles: $a=1.2$; and squares: $a = 2.2$, where $g_0(h) = \mathcal{N}(a) h^a e^{-h/b}$ (see text). The dashed vertical line is $\widehat{\Phi}$ for the final time-averaged $g(h)$.}
\label{fig:relaxation}
\end{figure}

We calculate $\Lambda_R$ as follows. We start with different initial conditions $g_0(h) = \mathcal{N}(a) h^a e^{-h/b}$ varying $a$ and $b$ to obtain different values of $\Phi$. We solve 
\be
\frac{\partial g}{\partial t} = \frac{\partial}{\partial h} \left(\phi g\right) + \frac{\partial^2}{\partial h^2}\left(k_2 g\right),
\label{eqn:fpt2_app2}
\ee
which is equation 11 of the main document, with $f$ given by 
\be
\begin{split}
f &= \frac{1}{\rho_i \, L_i \, f_0} \left[-\left(1-\alpha\right) F_S + F_0 + \sigma_T T  - \Delta F_0 - F_B \right] \\
&- \frac{1}{f_0} \nu_0 h,
\label{eqn:ew09_app3}
\end{split}
\ee
which is equation 18 of the main document, and compute the time it takes for $g(h,t)$ to reach a stationary state. This is done for $a = 1.2$ and $2.2$, and $b = [0.1, 1.2]$, giving a range of $\Phi$. Figure \ref{fig:relaxation} shows $\Lambda_R$ as a function of the thick-ice fraction ($\widehat{\Phi} = 1- \Phi$) of $g_0(h)$, and the dashed vertical line is $\widehat{\Phi}$ for the final time-averaged $g(h)$, denoted by $\widehat{\Phi}_f$. The two curves, representing different values of $a$, display similar behavior. This shows that $\Lambda_R$ is a function of $\widehat{\Phi}$ to leading order. When $\widehat{\Phi} < \widehat{\Phi}_f$, so that the thick-ice fraction of the initial condition is less than that of the stationary state, $\Lambda_R$ varies between 4 and 8 years.  Therefore, there is a range of of thin ice fractions, $\Phi \approx 0.3 - 0.6$ for which the relaxation time scale is approximately 50\% that of thermodynamic only models, viz., $\sim$ 4 years rather than $\sim$ 8 years, where there is an efficient mechanical redistribution of ice thickness, which is faster than solely thermodynamic time scales. 
However, when $\widehat{\Phi} > \widehat{\Phi}_f$, $\Lambda_R$ is a non-decreasing function of $\widehat{\Phi}$, and it is increasingly difficult for both thermal and mechanical processes to drive the system to the new state.

\section*{Appendix 4: Stability analysis of the Crank-Nicolson scheme for the advection-diffusion equation}

To study the stability properties of the Crank-Nicolson scheme when applied to an advection-diffusion equation with constant transport coefficients, we perform a von Neumann analysis.

Consider the following advection-diffusion equation
\be
\frac{\partial u}{\partial t} = V \frac{\partial u}{\partial x} + D \frac{\partial^2 u}{\partial x^2}, 
\ee
where $u$ is some quantity being transported, $V$ is the advection speed, and $D$ is the diffusivity. Using central differences for the spatial derivatives and semi-implicit C-N for time integration, we have
\begin{widetext}
\be
\frac{u_j^{n+1} - u_j^n}{\Delta t} = \frac{V}{2} \left\{\left[\frac{u_{j+1}^{n+1} - u_{j-1}^{n+1}}{2 \Delta x}\right] + \left[\frac{u_{j+1}^{n} - u_{j-1}^{n}}{2 \Delta x}\right] \right\}
+ \frac{D}{2} \left[\frac{u_{j+1}^{n+1} - 2 u_j^{n+1} + u_{j-1}^{n+1}}{(\Delta x)^2}\right] 
+  \frac{D}{2} \left[\frac{u_{j+1}^{n} - 2 u_j^{n} + u_{j-1}^{n}}{(\Delta x)^2}\right].
\label{eqn:fpt_cn}
\ee
\end{widetext}
On rearrangement, we find the following
\begin{widetext}
\be
u_j^{n+1} - u_j^n = p_1 \left(u_{j+1}^{n+1} - u_{j-1}^{n+1} + u_{j+1}^n - u_{j-1}^n\right) 
+ p_2 \left( u_{j+1}^{n+1} - 2 u_j^{n+1} + u_{j-1}^{n+1} + u_{j+1}^{n} - 2 u_j^{n} + u_{j-1}^{n} \right),
\label{eqn:fpt_cn_2}
\ee
\end{widetext}
where $p_1 = \frac{V \Delta t}{4 \Delta x}$ and $p_2 = \frac{D \Delta t}{2 \left(\Delta x\right)^2}$. Assuming wave-like solutions we have
\begin{widetext}
\be
u_j^n = \widehat{u}_n \, e^{i \, j \, k \, \Delta x}; \, u_j^{n+1} = \widehat{u}_{n+1} \, e^{i \, j \, k \, \Delta x}; \, u_{j+1}^{n} = \widehat{u}_{n} \, e^{i \, (j+1) \, k \, \Delta x}, \, \text{etc.}
\ee
\end{widetext}
where $i = \sqrt{-1}$ and $k$ is the wavenumber. Using these in equation \ref{eqn:fpt_cn_2} and after some algebra we obtain:
\be
G = \frac{1 + 2 \, i \, p_1 \, \sin\left(k \, \Delta x\right) - 4 \, p_2 \, \sin^2\left( \frac{k \, \Delta x}{2}\right)}{1 - 2 \, i \, p_1 \, \sin\left(k \, \Delta x\right) + 4 \, p_2 \, \sin^2\left( \frac{k \, \Delta x}{2}\right)} = \frac{a + i \, b}{c - i \, b},
\ee
where $G = \widehat{u}_{n+1}/\widehat{u}_n$ is the amplification factor, $a = 1 - 4 \, p_2 \, \sin^2\left( \frac{k \, \Delta x}{2}\right)$, $b = 2 \, p_1 \, \sin\left(k \, \Delta x\right)$, and $c =  1 + 4 \, p_2 \, \sin^2\left( \frac{k \, \Delta x}{2}\right)$. The squared amplitude of $G$ is
\be
|G|^2 = \frac{\left(ac - b^2\right)^2 + b^2 (a+c)^2}{\left(b^2 + c^2\right)^2} = \frac{\left(ac - b^2\right)^2 + 4 b^2}{\left(b^2 + c^2\right)^2}.
\label{eqn:neumann}
\ee

For stability, we must have $|G| \le 1$. Assuming $a = O(1)$ and $c = O(1)$, we consider the following cases:
\begin{enumerate}
\item When $b \ll 1$, we have
\be
|G|^2 \approx \frac{a^2}{c^2} = \left[\frac{1 - 4 \, p_2 \, \sin^2\left( \frac{k \, \Delta x}{2}\right)}{1 + 4 \, p_2 \, \sin^2\left( \frac{k \, \Delta x}{2}\right)}\right]^2.
\ee
This is the amplification factor for the pure diffusion equation for C-N scheme, and implies unconditional stability as $|G| \le 1$ for all cases.

\item When $b \gg 1$, we have
\be
|G|^2 \approx 1 + \frac{4}{b^2},
\ee
which implies numerical instability. 

\end{enumerate}

To ensure numerical stability in our simulations, $\Delta t$ and $\Delta h$ are chosen such that $|p_1| \le 0.4$ throughout the year.
 
\bibliography{g(h)_arxiv}

\end{document}